\documentclass[aps,prd,twocolumn,nofootinbib,superscriptaddress]{revtex4}

\usepackage{amssymb,graphics,graphicx, epsfig}
\usepackage{graphicx}
\usepackage{amsfonts}
\usepackage{amssymb}
\usepackage{dsfont}
\usepackage{bm}
\newcommand{\beqn}{\begin{eqnarray}}
\newcommand{\eeqn}{\end{eqnarray}}
\newcommand{\be}{\begin{equation}}
\newcommand{\ee}{\end{equation}}

\newcommand{\st}{{Stueckelberg~}}

\newcommand{\mathsym}[1]{{}}

\def\beq{\begin{equation}}
\def\eeq{\end{equation}}
\def\beqn{\begin{eqnarray}}
\def\eeqn{\end{eqnarray}}
\def\st{Stueckelberg\ }

\def\s1{$s_{\alpha}$}
\def\s2{$s_{\gamma}$}
\def\s3{$s_{\delta}$}
\def\c1{$c_{\alpha}$}
\def\c2{$c_{\gamma}$}
\def\c3{$c_{\delta}$}
\def\s{Stueckelberg~}

\begin{document}
\title{The Stueckelberg  $Z'$ Extension with Kinetic Mixing  \\ and  Milli-Charged Dark Matter from the Hidden Sector  }
\author{Daniel Feldman}
\email{feldman.da@neu.edu}
\author{Zuowei Liu }
\email{liu.zu@neu.edu }
\author{Pran Nath}
\email{nath@lepton.neu.edu} \affiliation{Department of Physics,
Northeastern University,
 Boston, MA 02115, USA \\ 
\rm V1: February 12, 2007 ; Published: June 1, 2007, Phys. Rev. D
75, 115001 2007  }
\begin{abstract}

 An analysis is given of  the Stueckelberg extension of  the Standard Model
with a hidden sector gauge group $U(1)_X$ where the mass growth for
the extra  gauge boson occurs via the  Stueckelberg mechanism, and
where the kinetic mixing in the $U(1)_X\times U(1)_Y$ sector is
included. Such a kinetic mixing is generic in a broad class of
supergravity and string models. We carry out a detailed global fit
of the model with the precision LEP data on and off the $Z$ pole,
with $\chi^2$ within 1\% of the $\chi^2$ of the Standard Model fit.
Further, it is shown that in the absence of matter in  the hidden
sector, there is a single effective parameter that controls the
deviations from the Standard Model predictions, and the  dependence
on the kinetic mixing emerges only when matter in the hidden sector
is included. An analysis is also given of milli-charged  dark matter
arising from the hidden sector, where it is shown that such dark
matter from the Stueckelberg extension can satisfy WMAP-3 data while
allowing for  a sharp $Z'$ resonance which can be detected at the
Tevatron and at the LHC via a dilepton signal generated by the
Drell-Yan process.

\end{abstract}
\maketitle
\section{Introduction}
Recent works on the Stueckelberg extension of the SM  \cite{kn1}
(StSM) and of the MSSM \cite{kn2,kn3} (StMSSM), have shown
consistency with the LEP data while allowing for the possibility of
a narrow  resonance   which could lie as low as just above the $Z$
mass \cite{kn1,kn2,kn3,fln1,fln2}. Similar phenomena regarding a
narrow resonance  are seen in other classes of models such as those
based on universal extra  dimensions \cite{UED}, models  with a
shadow sector \cite{holdom,Kumar:2006gm,Chang:2006fp}, and in the
models considered in Ref. \cite{Ferroglia:2006mj}. The possibility
of a narrow graviton resonance  also arises  in the RS model
\cite{Randall:1999ee,Davoudiasl:1999jd}. Thus the study of narrow
resonances is a topic of significant interest. In this work we focus
on the Stueckelberg extensions further.   Such models may be  a
field theory realization of string models arising from orientifolds
\cite{ghi,Ibanez:2001nd,Anastasopoulos:2006cz,recent_ST,Anastasopoulos:2007qm}
and their  discovery could be harbingers  of a new regime of physics
altogether. The supersymmetric  Stueckelberg  extensions have the
possibility of generating a new type of neutral Majorana fermion
which is massive and extra-weakly interacting, and with R parity
conservation, a candidate for cold dark matter \cite{Feldman:2006wd}.
Indeed a detailed analysis shows that the
relic density in such models is consistent with the three year WMAP
data \cite{Spergel:2006hy}. Additionally, the Stueckelberg
extensions give rise to milli-charges for matter residing in the
hidden sector\cite{kn1,kn3}. A recent analysis \cite{kctc}
indicates that such matter can annihilate in sufficient amounts to
satisfy the relic density constraints from WMAP-3  for a broad $Z'$
resonance.

The main focus of  this work is an extension of the class of models
considered in Refs. \cite{kn1,kn2,kn3,fln1,fln2} by including
kinetic mixing between the two Abelian $U(1)$ gauge fields.
Specifically we consider  the extended electroweak sector with the
gauge groups $SU(2)_L\times U(1)_Y\times U(1)_X$ where the
Stueckelberg mechanism along with the spontaneous breaking in the
Higgs sector generates the vector boson mass, and a mixing in the
gauge kinetic energy of the $U(1)_X\times U(1)_Y$ sector is
included. Such kinetic mixings can arise in a variety of ways
\cite{bkm,dkm} and enter prominently in models of the type
considered in Refs. \cite{Kumar:2006gm,Chang:2006fp}. The model
considered here encompasses the models of Refs.\cite{kn1,kn2,kn3,fln1,fln2}
which can be obtained in certain
limits of the model discussed here.   Inclusion of the kinetic
mixing in the Stueckelberg extension enhances significantly the
parameter space where new physics can exist consistent with the
stringent LEP, Tevatron, and WMAP constraints.  This parameter space
includes the possibility of a narrow
$Z'$ resonance  very distinct from the $Z'$ of the conventional models \cite{u1,d1,PhemZp}. \\

The outline of the rest of the paper is as follows: In Sec.(II) we
give a description of the \st extension with kinetic
mixing in the $U(1)_X\times U(1)_Y$ sector. In Sec.(III) we give a
detailed numerical analysis of the electroweak constraints from LEPI
and LEPII.  In Sec.(IV) an analysis of milli-charged dark matter
that arises from the hidden sector is given. Conclusions are given
in Sec.(V). Appendices  contain several mathematical details,
including an analysis regarding the origin of milli-charged matter.

\section{The Stueckelberg $Z' $ Extension with Kinetic Mixing}
In this section we discuss  the $U(1)_X$ Stueckelberg \cite{stueck}
extension of the Standard Model (SM) with gauge kinetic mixing
(StkSM). We assume that the quarks, leptons and the Higgs field of
the SM do not carry $U(1)_X$ quantum numbers, and the fields in the
hidden sector do not carry quantum numbers of the SM gauge group.
Thus the $U(1)_X$ sector is hidden except for the $U(1)_X\times
U(1)_Y$ mixings in the gauge vector boson sector as given by the
effective Lagrangian
\begin{eqnarray}
\mathcal{L}_{\rm StkSM} &=& \mathcal{L}_{\rm SM} +\Delta  \mathcal{ L}, \nonumber\\
\Delta\mathcal{L} & \supset &  -\frac{1}{4} C_{\mu \nu}C^{\mu \nu}
- \frac{\delta}{2}C_{\mu\nu}B^{\mu \nu} \nonumber\\
&& -\frac{1}{2}(\partial_{\mu}\sigma+ M_1 C_{\mu}+M_2 B_{\mu})^2
+ g_{X}J^{\mu}_{X}C_{\mu}\nonumber.\\
\label{stksm}
\end{eqnarray}
Here  $B_{\mu}$ is the $U(1)_Y$ gauge field,  $C_{\mu}$ is the
$U(1)_X$ gauge field with coupling strength  $g_X$ to hidden matter
through the source $J_X$, $\sigma$ is the axion, $M_1$ and $M_2$ are
mass parameters, and $\delta$  is the kinetic mixing parameter.  The
$\sigma$ field is charged under both $U(1)_X$ and $U(1)_Y$ and the
Lagrangian of Eq. (1) is invariant under the $U(1)_X\times U(1)_Y$
gauge transformations \beqn
\delta B_{\mu}= \partial_{\mu} \lambda_X, ~\delta C_{\mu}=0, ~~\delta \sigma =-M_2\lambda_X\nonumber\\
\delta B_{\mu}=0, ~\delta C_{\mu}= \partial_{\mu} \lambda_Y, ~
\delta \sigma =-M_1\lambda_Y. \eeqn The above model has a
non-diagonal kinetic mixing matrix  ($\mathcal{K}$)  and a
non-diagonal mass matrix  (${M}^2_{\rm St}$) and  in the unitary
gauge \cite{kn1} in the basis $V^T = (C, B, A^3)$
\begin{equation}
 \mathcal{K}= \left[\matrix{
 1 & \delta &0  \cr
    \delta & 1 &0  \cr
    0 & 0 & 1
}\right],\end{equation}\begin{equation} {M}_{{ \rm St}}^2=
 \left[\matrix{
    M_1^2             & M_1M_2                                & 0 \cr
    M_1M_2         &M_2^2 + \frac{1}{4}v^2g_Y^2  & -\frac{1}{4}v^2g_2g_Y \cr
      0                   & -\frac{1}{4}v^2g_2g_Y            & \frac{1}{4}v^2g_2^2
}\right].
\label{stmass}
 \end{equation}
 A simultaneous diagonalization of the kinetic energy and of the mass matrix can
 be obtained by a transformation $T=KR$, which is a combination of a $GL(3)$ transformation ($K$)
 and an orthogonal transformation ($R$). This allows one to work in the diagonal basis,
 denoted by $E$ where $E^T=(Z',Z,A_{\gamma})$, through the  transformation  $V =(K R) E$,
where the matrix $K$ which diagonalizes  the kinetic terms has the form
\begin{equation}
 K = \left[\matrix{
 C_{\delta} & 0 &0 \cr
    -S_{\delta} & 1 &0 \cr
    0 & 0 & 1
}\right],\hspace{.15 cm}
C_{\delta}=\frac{1}{\sqrt{1-\delta^2}},\hspace{.15 cm}
S_{\delta}=\delta C_{\delta}.
\end{equation}
The matrix $R$ is then defined by the diagonalization of the mass
matrix \be M^2_D= R^T (K^T M^2_{\rm St} K) R. \ee The model of
Eq.(1)  involves three parameters: $M_1, M_2, \delta$ or alternately
$M_1, \epsilon, \delta$ where $\epsilon= M_2/M_1$. Expressing the
transformation $T=KR$ in terms of the matrix elements, one has \be
\begin{array}{l}\label{rotationmatrix}
 T_{1j} =  C_{\delta}R_{1j}, \\\nonumber
 T_{2j} =  R_{2j}-S_{\delta}R_{1j}, \\\nonumber
 T_{3j}=   R_{3j}.\nonumber
 \end{array}
\ee The neutral current interaction with the visible sector fermions
is given by \be {\mathcal L }_{NC} = -\frac{1}{i} \sum\limits_{f } [
{\bar f}_L D_{\mu} \gamma^{\mu} f_L + (L \rightarrow R)], \ee where
$D_{\mu}$ is the covariant derivative with respect to $SU(2)_L\times
U(1)_Y\times U(1)_X$ gauge group except that, as mentioned  in the
beginning, we assume that  the visible sector matter, i.e., quarks,
leptons and the Higgs,  are not charged  under $U(1)_X$.  Thus the
covariant derivative  includes  only the $SU(2)_L$ gauge coupling
$g_2$ and $U(1)_Y$ gauge coupling $g_Y$. The diagonalization also
leads to the following relation for the  electronic charge
\begin{equation}
\frac{1}{e^2}=\frac{1}{g_2^2}+\frac{1-2\epsilon\delta+\epsilon^2}{g_Y^2}.
\label{e}
\end{equation}
Thus  $g_Y$ is related to $g_Y^{SM}$ by
\be g_{Y}= \gamma \sqrt{1+\epsilon^2 -2 \delta
\epsilon},\hspace{.5cm} \gamma \equiv g^{SM}_Y,
\ee
and one may  write the neutral current interaction so that
\begin{widetext}\begin{eqnarray}\label{ncvs} {\cal L}_{NC} =
\frac{\sqrt{g^2_2+\gamma^2}}{2}\bar f \gamma^\mu \left[ (v'_{f}
-\gamma_5 a'_{f})Z'_\mu +( v_{f} -\gamma_5 a_{f})Z_\mu
 \right ] f + e \bar f \gamma^\mu Q_f  A_\mu f,
\end{eqnarray}
\end{widetext} where \be
\begin{array}{l}
v_f =\frac{1}{\sqrt{g^2_2+\gamma^2}}[ (g_2 T_{3 2} - g_Y T_{2 2})T^{3 }_f + 2  g_Y T_{2 2}Q_f ], \\
a_f =\frac{1}{\sqrt{g^2_2+\gamma^2}}[ (g_2 T_{3 2} - g_Y T_{2 2})T^{3 }_f   ], \\
v'_f  =\frac{1}{\sqrt{g^2_2+\gamma^2}}[ (g_2 T_{3 1} - g_Y T_{2 1})T^{3 }_f  + 2  g_Y T_{2 1}Q_f ], \\
a'_f =\frac{1}{\sqrt{g^2_2+\gamma^2}}[ (g_2 T_{3 1} - g_Y T_{2 1})T^{3 }_f  ], \\
 \end{array}
\ee
and where, as usual, $Q_f= T^3_f+ Y_f/2$.
We note that the mass and kinetic mixing parameters  enter not only  through $T_{ij}$ but also
through $g_Y$ via the constraint of Eq.(\ref{e}).

\section{Constraints from \newline electroweak data}
We discuss now the constraints on the model  with both mass mixing and kinetic mixing
from the precision electroweak data.  We start by assuming
 that the hidden sector does not contain matter,  and
the case when matter is included in the  hidden sector is discussed
in Sec.(IV) and in Appendix B. To obtain the allowed range of
$\epsilon$ and $\delta$, we follow the same approach as  in Ref.
\cite{Nath:1999fs,fln1,fln2}.  The first constraint comes from the
comparison of the one sigma error in the prediction of the $Z$ boson
mass in the Standard Model and a comparison of this result with
experiment, leads to an error corridor,  $\delta M_Z\sim  37$ MeV,
where one can accommodate new physics. However, the more stringent
constraint comes from fits to the high precision LEP data on  the
branching ratios of the $Z$ decay and from the various asymmetries
at the  $Z$ pole, when one demands that the $\chi^2$ fits of StkSM
are within 1\% of that of the Standard Model. We will refer to this
as the LEPI 1\% constraint in the rest of the paper.
   Details of the method employed for the electroweak fits can be found in the analysis of Ref. \cite{fln1,fln2},
   and here we present just the results in the extended model with kinetic mixing.\\

\begin{figure}[t]
\vspace*{.2in} \hspace*{-.2in} \centering
\includegraphics[width=8 cm,height=8 cm]{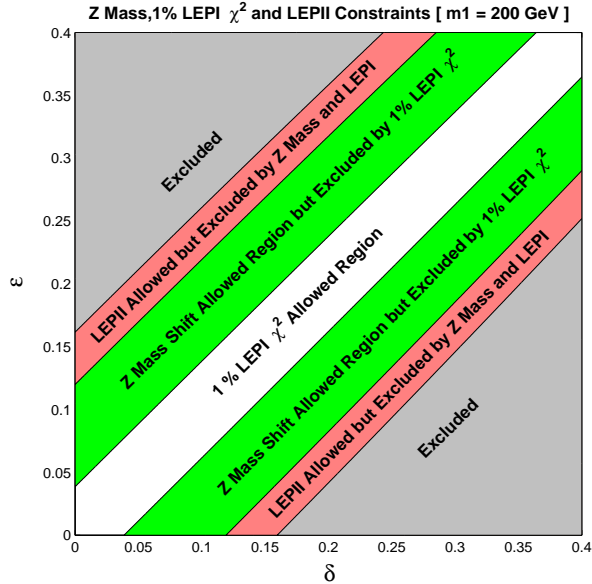}
\caption{An analysis of $\chi^2$ in StkSM model in the
$\epsilon-\delta$ plane.   The center white region is where $\chi^2$
of the StkSM model is within 1\% of the SM fits. Along the line
$\epsilon=\delta$ the $Z'$ decouples and the model gives the same
$\chi^2$ fit to data as in the  SM (see also Appendix A).}
\label{200}
\end{figure}

\begin{figure}[t]
\vspace*{.4in} \hspace*{-.2in} \centering
\includegraphics[width=8 cm,height=7 cm]{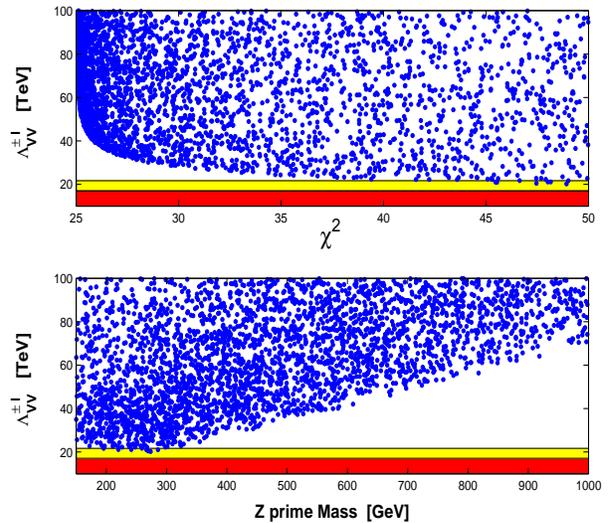}
\caption{The plots give an analysis of the LEPII constraint. The
upper plot, which has a $Z'$ mass range of $ .2 - 1$ TeV, shows the
relevant LEPII contact interaction parameter $\Lambda^{\pm l}_{VV}$
as a function of $\chi^2$ for the 19 observable of Table(1),  where
a $\chi^2 \sim$ 25 is the SM fit
 as given in
 Table(1), and where the (yellow,red) shaded
regions correspond to $\Lambda^{\pm l}_{VV}$ = (21.7, 17.1) TeV
\cite{Leptwo}. The lower plot for $\chi^2$ in the range (25-50)
gives  $\Lambda^{\pm l}_{VV}$ as a function of the $Z$ prime mass.}
\label{Contact2}
\end{figure}

Table 1 gives a fit to the LEP data for  a specific point in the \st
parameter space with
 $\epsilon = .06$, $\delta = .03$,  and $ M_{Z'} \approx M_1 = 200$ GeV.
  In the analysis we have taken into account the constraint between $g_Y$ and $g_Y^{SM}$ and the
 inclusion of this constraint  improves  the electroweak  fits over that of previous analyses for the case $\delta= 0$ \cite{fln1,fln2}.
 Thus the analysis of Table 1 shows that in the StkSM  one finds $\chi^2$ fits which are at the
 same level as in the SM.
 An analysis of $\chi^2$ in the LEPI fits in the $\epsilon-\delta$ parameter space  is given
    in Fig.(\ref{200}).
    Specifically   Fig.(\ref{200}) shows that a large region of the parameter
    space can satisfy the LEPI 1\% constraint.
  A striking  aspect of the analysis of   Fig.(\ref{200}) is that this constraint is
  satisfied even though $\epsilon$ and $\delta$ can get significantly large,
  as long as $(\epsilon -\delta)$ is small.  The physics of this is explained in Appendix A
  where it  is shown that in the absence of matter in the hidden sector,
    there is  only one effective parameter, $\bar\epsilon$,
   that enters  the analysis of electroweak physics.
\begin{table*}[t]
\caption{\label{tab:table2} Fits to 19 $Z$ pole observables. Column
2 is given by the PDG \cite{pdg}, while the data in column 3 is from
the SM Fit of the LEP EWWG \cite{:2005em}. The column labeled St Fit
is an analysis for the input $\epsilon =0.06$, $ \delta= 0.03$, and
$M_1 = 200$ GeV. In the last column PULL is defined by $(\rm
Experiment  - \rm FIT)/\Delta$,  and
 $\chi^2 = \sum{\small{\rm{PULL}}^2}$.
}
\begin{ruledtabular}
\begin{tabular}{ccccccccc}
Quantity                 & Experiment $\pm \Delta$     &LEP FIT          &St FIT        &LEP PULL     & St PULL       \\
\hline
$\Gamma_Z$ [GeV]           & 2.4952  $\pm$ 0.0023      &2.4956       &2.4956    & -0.17   & -0.17     \\
$\sigma_{\rm had}$ [nb]        & 41.541  $\pm$ 0.037       &41.476       &41.469    & 1.76   &1.95    \\
$R_e$                     & 20.804  $\pm$ 0.050        &20.744       &20.750     &  1.20      & 1.08   \\
$R_\mu$                    & 20.785  $\pm$ 0.033       &20.745       &20.750    &  1.21   & 1.06   \\
$R_\tau$                   & 20.764  $\pm$ 0.045       &20.792       &20.796     &  -0.62  & -0.71  \\
$R_b$                      & 0.21643 $\pm$ 0.00072     &0.21583     &0.21576     & 0.83    &  0.93 \\
$R_c$                      & 0.1686 $\pm$ 0.0047       &0.17225      & 0.17111  & -0.78    &  -0.53  \\
$A^{(0,e)}_{FB}$           & 0.0145 $\pm$ 0.0025       &0.01627       &0.01633  & -0.71   & -0.73\\
$A^{(0,\mu)}_{FB}$         & 0.0169  $\pm$ 0.0013      &0.01627       & 0.01633  & 0.48   &  0.44   \\
$A^{(0,\tau)}_{FB}$        & 0.0188  $\pm$ 0.0017      &0.01627       & 0.01633 & 1.49   &  1.45\\
$A^{(0,b)}_{FB}$           & 0.0991  $\pm$ 0.0016      &0.10324       & 0.10344  & -2.59   &  -2.71 \\
$A^{(0,c)}_{FB}$           & 0.0708  $\pm$ 0.0035      &0.07378       &0.07394   & -0.85   & -0.90\\
$A^{(0,s)}_{FB}$           & 0.098  $\pm$ 0.011        &0.10335       &0.10355    & -0.49   &  -0.50 \\
$A_e$                      & 0.1515 $\pm$ 0.0019       &0.1473        &0.1476  &  2.21  & 2.05 \\
$A_\mu$                    & 0.142  $\pm$ 0.015        &0.1473        &0.1476    & -0.35   &  -0.37 \\
$A_\tau$                   & 0.143  $\pm$ 0.004        &0.1473        &0.1476    & -1.08   &  -1.15\\
$A_b$                      & 0.923  $\pm$ 0.020        &0.93462       &0.93464    & -0.58   &   -0.58 \\
$A_c$                      & 0.671  $\pm$ 0.027        &0.66798       &0.66812    & 0.11   &   0.11 \\
$A_s$                      & 0.895 $\pm$ 0.091         &0.93569       &0.93571    & -0.45   & -0.45 \\
                                                                & & & & $\chi^2=$25.0&
                                                              $\chi^2=$25.2
\end{tabular}
\end{ruledtabular}
 \end{table*}

    We discuss next the LEPII contraints. These consraints are
typically characterized by the parameter of contact interaction
$\Lambda$, and  the
 LEPII group
 finds that $\Lambda_{VV}> (21.7, 17.1)$ TeV \cite{Leptwo}
   to be the most constraining.
The StkSM model predicts the theoretical value of $\Lambda_{VV}$
through the following formula
\begin{equation}
\Lambda_{VV}=\frac{M_{Z'}}{M_Z}\sqrt{\frac{4\pi}{\sqrt{2}G_Fv_e^{'2}}}.
\label{lambdavv}
\end{equation}
 A numerical analysis of the LEPII constraints is given in Fig.(\ref{200})
 and  Fig.(\ref{Contact2}).
  The analysis  of Fig.(\ref{200})
 exhibits that the LEPI 1\% constraint is more  stringent than the LEPII constraint, and
 thus the LEPII constraint is
 automatically satisfied once the LEPI 1\% constraint is satisfied.
 This result is supported by the analysis of
    Fig.(\ref{Contact2}) which  shows  that the value
 of $\Lambda_{VV}$ predicted by the model in the parameter space consistent with
the LEPI 1\% constraint is  significantly larger than the lower
limit of  the LEPII constraint.
 The blue points that enter the
shaded regions are eliminated by the LEPII constraint.  However,
these points also correspond to large $\chi^2$ fits to the LEPI analysis
and are eliminated by LEPI 1\% constraint as well.
Thus, for a narrow $Z'$, the LEPI 1\% constraint is
stronger than the LEPII constraint.

     \section{Milli-charged dark matter from the hidden sector}
     In the previous section we did not include matter in the hidden sector which is defined as matter which is neutral
     under the SM gauge group but carries $U(1)_X$ quantum numbers and thus couples only to $C_{\mu}$.
     The kinetic and mass mixings in the $U(1)_X\times U(1)_Y$ sectors typically generate milli-charges for such matter.
        The  conditions for the origin of milli-charges  arising  from such mixings
        are discussed in Appendix B, where
     simple examples are  worked out to explain the constraints that lead to the appearance of  such  charges.
         Milli-charges have been examined in many works both theoretically and
     experimentally \cite{holdom,Goldberg:1986nk,Golowich:1986tj,Mohapatra:1990vq,Davidson:1993sj,Foot:1989fh,Caldwell:1988su, Dobroliubov:1989mr,Foot:1994bx,Davidson:2000hf,Perl:2001xi,Prinz:1998ua,Dubovsky:2003yn,Masso:2006gc,Abel:2003ue,Abel:2006qt,Ahlers:2006iz,Badertscher:2006fm}. Most of these analyses are in the context of kinetic mixing model of \cite{holdom}.
      Here we  consider the milli-charged matter in the hidden sector
     within the context of the \st extension of the SM
     with both mass and kinetic mixing.
     If milli-charged
matter exists then both the $Z$ and the $Z'$ can decay into it if
kinematically allowed to do so.  For the mass scales we  investigate
the milli-charge particle has a mass larger than $M_Z/2$. In this
case all of the electroweak constraints discussed in Sec.III are
unaffected. Further, the $Z$ prime can decay into the millicharged
matter if the mass of the hidden matter is less than $M_{Z'}/2$.
Such decays increase the $Z'$ width and thus decrease the branching
ratios of the $Z'$ decay into the visible sector which depletes the
dilepton signal in the Drell-Yan process. A relatively strong
dilepton signal manifests in the  analysis of Refs.
\cite{kn3,fln1,fln2} where the $Z'$ decays into the hidden sector
were taken to be comparable to the $Z'$ decays into the visible
sector, i.e., $\Gamma_{Z'}^{\rm hid}\sim \Gamma_{Z'}^{\rm vis}$.
This constraint then  leads to a sharp $Z'$ resonance, but the decay
of the hidden sector matter via the $Z'$ pole is not strong enough
to annihilate the hidden sector milli-charged dark matter in
sufficient amounts to be consistent with experiment, unless extreme
fine tuning is used (we return to this issue at the end of this
section).
\\
\begin{figure}[t]
\vspace*{.2in} \hspace*{-.2in} \centering
\includegraphics[width=8 cm,height=6 cm]{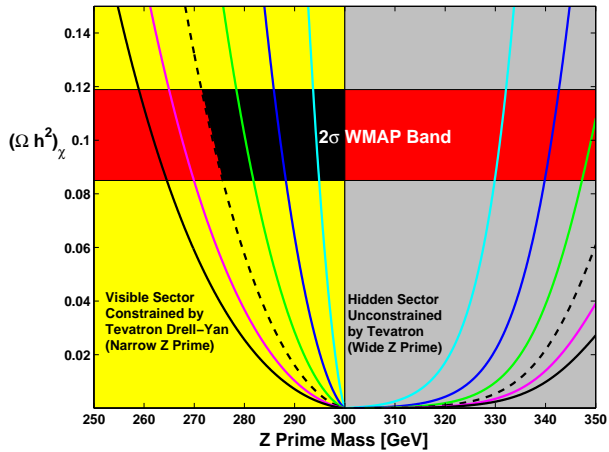}
\caption{An analysis of  the relic density of milli-charged
particles  arising in the StSM  $Z'$ model from the hidden sector
for the case $\delta =0$, $M_{\chi} = $ 150 GeV, $\epsilon =
(.01-.06)$ with .01 for the innermost curve and moving outward in
steps of .01.  The (yellow, grey) regions ($M_{Z'}<2M_{\chi}$,
$M_{Z'}>2M_{\chi}$) correspond to a (narrow, broad) $Z'$ resonance,
and the WMAP-3 relic density constraints are satisfied for  both a
broad $Z'$ resonance and a narrow $Z'$ resonance as exhibited by the
 2$\sigma$  WMAP-3  red and black bands. The region of  narrow  $Z'$ resonance is constrained
 by the LEP and Tevatron data. The region in the  $2\sigma$ WMAP-3 band can be probed
 via a dilepton signal as shown in Fig.(\ref{relic4}).  The red band to the left is excluded
 by the CDF 95\% C.L. \cite{CDF} data while the black band is consistent or on the edge thereof, with all constraints
 which can produce an observable dilepton signal. }
\label{relic3}
\end{figure}

\begin{figure}[t]
\vspace*{.2in} \hspace*{-.2in} \centering
\includegraphics[width=8 cm,height=6 cm]{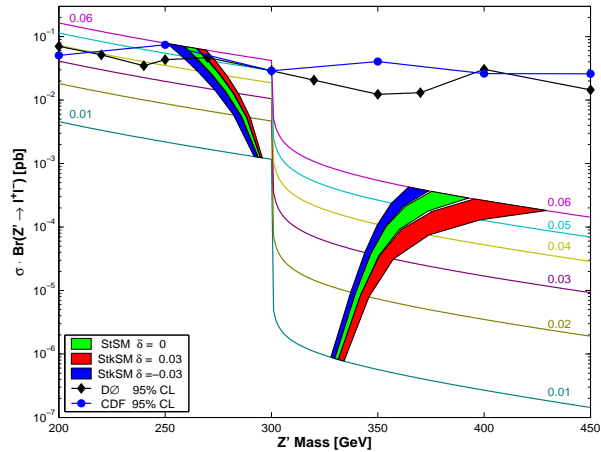}
\caption{An exhibition of the dilepton signal $\sigma \cdot
Br(Z'\rightarrow l^+l^-)$ at the Tevatron consistent with the WMAP-3
relic density constraint  as a function of  $M_{Z'}$ when
$2M_{\chi}=300$ GeV. The curves in ascending order are for values of
$\bar{\epsilon}$ in the range $(0.01-0.06)$ in steps of 0.01. The
dilepton signal has a dramatic fall as $M_{Z'}$ crosses the point
$2M_{\chi}=300$ GeV where the $Z'$ decay into the hidden sector
fermions is kinematically allowed, widening enormously the $Z'$
decay width. The green shaded regions are where the WMAP-3 relic
density constraints are satisfied for the case when there is no
kinetic mixing. Red and blue regions are for the case when kinetic
mixing is included. The D\O\ data set \cite{Abazov:2005pi} was
collected in the search for narrow resonances (RS) and is a stronger
constraint to apply on this model than the recent  CDF \cite{CDF}
data which put constraints on the parameter space when the $Z'$ can
decay only into matter in the visible sector. }
 \label{relic4}
\end{figure}

     The recent work of  Ref. \cite{kctc}  has carried out an explicit analysis
     of putting a pair of Dirac fermions in the hidden sector, and made the interesting observation that
    for values $g_XQ_X \leq O(1)$ the decay width of $Z'$ into the hidden sector Dirac fermions ($\chi$)  can be
    of GeV size, and consequently the hidden matter can annihilate in sufficient amounts to satisfy
    the relic density.
    We have carried out a similar analysis  using the  thermal
    averaging procedure in the computation of the relic density as  described  in Appendix C.
     Our conclusions  are in agreement  with the  analysis of Ref. \cite{kctc} in the region of the
    parameter space  investigated in Ref. \cite{kctc} when no kinetic mixing is assumed in the absence
    of thermal averaging.
    In our work we  take the  kinetic mixing into account in the analysis of the relic density.
    We also  make a further  observation that there exists a  significant region of the parameter space where
    it is possible to satisfy the relic density constraints
    and still have a narrow $Z'$ resonance which can be detected at the Tevatron and at the LHC
     using  the dilepton signal via  a Drell-Yan process. \\

\begin{figure}[b]
\vspace*{.2in} \hspace*{-.2in} \centering
\includegraphics[width=8 cm,height=6 cm]{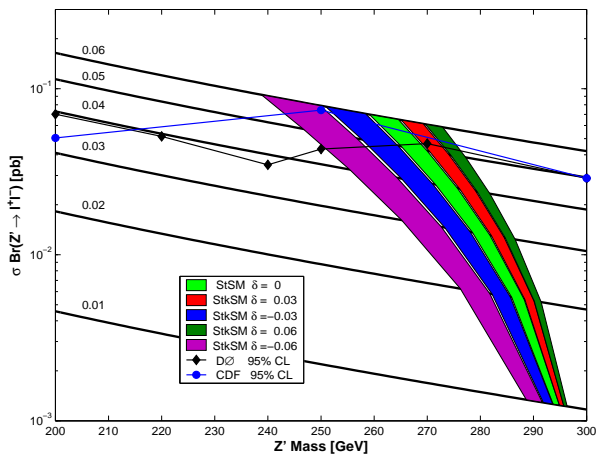}
\caption{Same as  Fig.(\ref{relic4}) except that only the mass
region with a detectable dilepton signal at the Tevatron is
exhibited but additional $\delta$ values are included in the
analysis. The plots show that the region allowed by WMAP-3
constraints moves to the right for positive $\delta$ and to the left
for negative $\delta$. This phenomenon is explained in Appendix D.
The  inclusion of kinetic mixing is seen to enlarge the parameter
space where the relic density constraints are satisfied and where an
observable dilepton signal at the Tevatron can occur. }
\label{relic4a}
\end{figure}

\begin{figure}[b]
\vspace*{.2in} \hspace*{-.2in} \centering
\includegraphics[width=8 cm,height=6 cm]{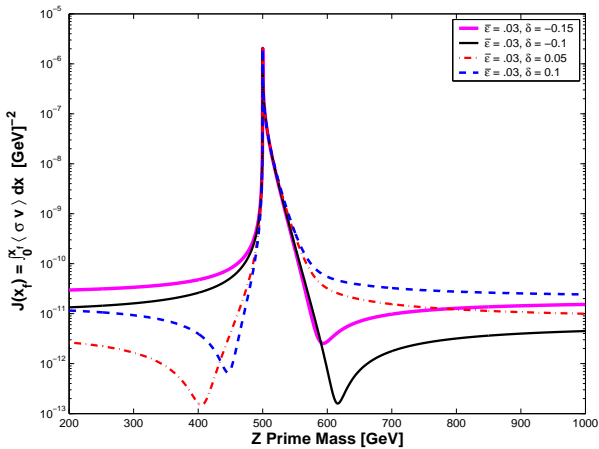}
\caption{ Illustration of the variation with $\delta$ of the thermally averaged cross
section for the annihilation of milli-charged particles  when
$M_{\chi} = 250 $ GeV and $\bar{\epsilon} =.03$. The generic effect of
$\delta$ is to modify the shape of $J(x_f)$ and the region where the
relic density constraints  can be satisfied is enlarged. Further
details are given in appendix D.
 }
\label{J2}
\end{figure}

     We give now further details of our relic density analysis.
      In the numerical analysis we will use  $g_X=g^{SM}_{Y}$ and $Q_X=1$ unless stated otherwise.
         We begin by considering the case
     when $\delta =0$ which is the StSM case. Fig.(\ref{relic3}) gives an analysis of the relic density as a function of
     $M_{Z'}$ for the case $M_{\chi}=150$ GeV, and $\epsilon$ in the range
     $(0.01-0.06)$.   Here one finds that the relic density is satisfied
     on two branches, one for $M_{Z'}>2 M_{\chi}$, and the other for $M_{Z'}<2 M_{\chi}$.
     In the region $M_{Z'}>2 M_{\chi}$  the relic density is satisfied over a  broad range of $Z'$ masses
     for appropriate $\epsilon$ values. In this region the decay width  of the $Z'$ is large,
     and thus the branching ratio into the visible sector is suppressed, which would make the
     dilepton signal from
     the $Z'$ decay difficult to observe. This is exhibited in Fig.(\ref{relic4})
     where one finds that the dilepton signal essentially disappears in the region  $M_{Z'}>2 M_{\chi}$.
      Next we  examine the region   $M_{Z'}<2 M_{\chi}$.
         Here one finds once again a satisfaction
    of the relic density even though the $Z'$ mass lies below $2M_{\chi}$.  The very sizable range
    of the $Z'$ mass over which the relic density can be satisfied arises due to thermal averaging over
    the $Z'$ pole.   As expected in this region the $Z'$ will appear as a sharp  narrow resonance
    since the decay of the $Z'$ into the hidden sector fermions is kinematically disallowed.
An analysis of the dilepton signal for this case is also given in
Fig.(\ref{relic4}), and a more detailed view is given in
Fig.(\ref{relic4a}), which shows that  the Drell-Yan signal $p
\bar{p}\to Z'\to e^+e^-$ is enormously enhanced for
$M_{Z'}<2M_{\chi}$. Thus we  have a  region here of the parameter
space where one will have a sharp resonance giving a visible
dilepton signal while at the same time producing milli-charge dark
matter consistent with WMAP-3.

      We carry out  a similar analysis for the case when the kinetic mixing parameter $\delta$ is non-vanishing.
   Fig.(\ref{J2}) gives an analysis of the effect of $\delta$ on the thermally averaged cross section $J(x_f)$,
   defined in Appendix C,  as a function of the $Z'$ mass. One finds  that $J(x_f)$ is affected in a significant
   way by the change in $\delta$ and the particulars of the modification due to $\delta$  are explained in Appendix D.
    In Fig.(\ref{relic5}) an analysis of relic density as a function of $M_{Z'}$  is given for $\bar \epsilon =0.04$
   and $\delta$ in the range $\delta = (.05-.25)$. The analysis shows the relic density exhibits   a strong $\delta$
   dependence. Quite significantly, the region in $M_{Z'}$   over which the relic density can fall in the
   WMAP-3 region is widened.  Once again, in the region $M_{Z'}>2 M_{\chi}$ the dilepton signal will be
   too dilute to be observable as shown in Fig.(\ref{relic5}). On the other hand, in the region
   $M_{Z'}<2 M_{\chi}$, the $Z'$ is a sharp narrow resonance since the $Z'$ decay into the hidden sector fermions
   is kinematically disallowed. Thus the dilepton signal here is strong as can be seen in Fig.(\ref{relic4}) and
   Fig.(\ref{relic4a})  and  should
   be observable with sufficient luminosity.

\begin{figure}[htb]
\vspace*{.2in} \hspace*{-.2in} \centering
\includegraphics[width=8 cm,height=6 cm]{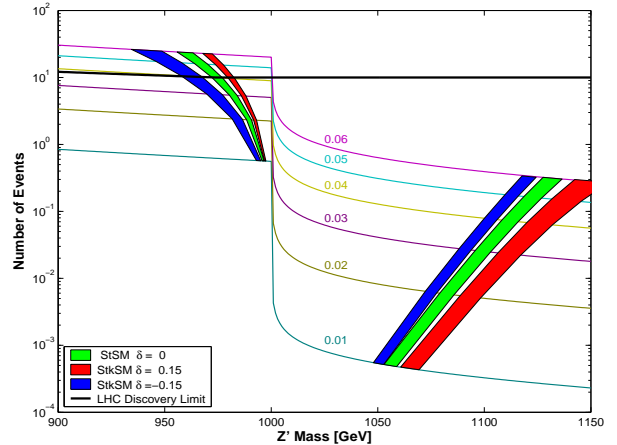}
\caption{ An exhibition of the dilepton signal as given by the
number of events for 10$fb^{-1}$ of integrated luminosity  at the
LHC consistent with the  WMAP-3 relic density constraint  as a
function of  $M_{Z'}$ when $M_{\chi}=500$ GeV. The curves in
ascending order are for values of $\bar{\epsilon}$ in the range
$(0.01-0.06)$ in steps of 0.01. Criteria set for the discovery limit
are discussed in the text. } \label{LHC}
\end{figure}

An analysis  for the case of the LHC in the Drell-Yan process $p p
\to Z'\to e^+e^-$, is given in Fig.(\ref{LHC}). Here it is shown
that the dilepton signal in the region $M_{Z'}<2M_{\chi}$
consistent with the WMAP-3 constraints would be discernible with
$10fb^{-1}$ of integrated luminosity for $\bar{\epsilon} \gtrsim
.04$. The dilepton signal has a dramatic fall as $M_{Z'}$ crosses
the point $M_{Z'}= 2M_{\chi}$  beyond which  the $Z'$ decay into the
hidden sector fermions is kinematically allowed, widening enormously
the $Z'$ decay width. The green shaded regions are where the WMAP-3
relic density constraints are satisfied for the case when there is
no kinetic mixing. Red and blue regions are for the case when
kinetic mixing is included. When computing the dilepton signal, a
20 GeV mass bin is used, and  a 50\% detector cut is assumed. The
thick solid line corresponds to the LHC discovery limit  with
$10fb^{-1}$ of integrated luminosity expected in the first year run,
and  the criteria used  for the discovery limit is $5\sqrt{N_{SM}}$
or 10 events, whichever is larger, where $N_{SM}$ is the number of
SM background events.    The analysis shows that data in the first
year running at the LHC should be able to constrain the parameter
space down to $\bar{\epsilon}\sim 0.04$ for $M_{\chi}=500$ GeV  when
$M_{Z'} < 1$ TeV.
\\

\begin{figure}[t]
\vspace*{.2in} \hspace*{-.2in}
\centering
\includegraphics[width=8 cm,height=6 cm]{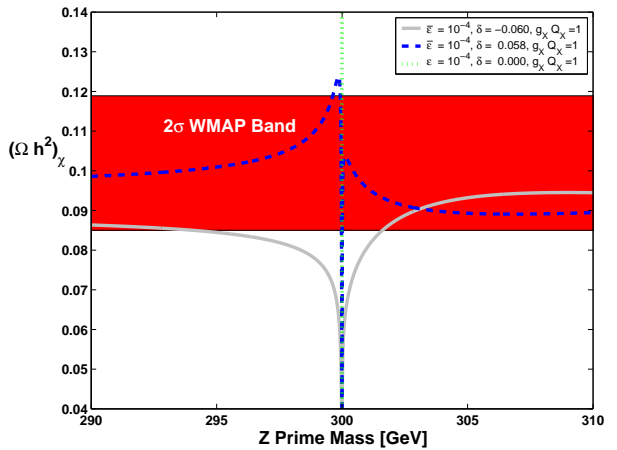}
\caption{An analysis of the  relic density of milli-charged
particles for the case of a very small $\bar{\epsilon}$.
 One finds that even for this case one can satisfy the WMAP-3 relic density
 constraints. In this case the hidden sector fermion mass is predicted
 to be essentially $M_{Z'}/2$.    While the  dilepton signal
 from the Drell-Yan process is too weak to be observable at the Tevatron,
 it is strong enough to be detectable at the LHC with  sufficient integrated
 luminosity \cite{fln2}.}
 \label{mini}
\end{figure}

An interesting issue concerns the question regarding how small $\bar
\epsilon$ can be  for WMAP-3 relic density constraints to be
satisfied. Fig.(\ref{mini}) addresses this question where an
analysis of the relic  density is given when $\bar
\epsilon=10^{-4}$.   One finds satisfaction of the relic density in
this case, and even smaller $\bar\epsilon$ were  found admissible.
Further, while the dilepton  signal at the Tevatron in this case
will be suppressed, it could still be visible at the LHC with
sufficient luminosity.
\\
\\
\\
\\

Finally, we note that within the context of the \st model it is
possible to place indirect limits on the milli-charge coupling of
hidden sector fermion with the photon from the Tevatron data. An
analysis of the limits on the \st mixing parameter $\epsilon$ was
presented in \cite{fln1} for the case $\delta=0$. For this case, the
milli-charge $Q_{milli}$, where $Q_{milli} e $ is the coupling of
the photon with the hidden sector fermions (see Appendix D) is
determined to be: $Q_{milli}\approx \epsilon $. Thus one may
directly translate the limits obtained in \cite{fln1} to limits on
the milli-charged coupling of the hidden fermion with the photon. In
the context of the present analysis, the cross-section predictions
given here, (as for example, in Fig.(\ref{relic4a})) with their
overlapping WMAP-3 bands, shows this explicitly.


\begin{figure}[t]
\vspace*{.2in} \hspace*{-.2in} \centering
\includegraphics[width=8 cm,height=6 cm]{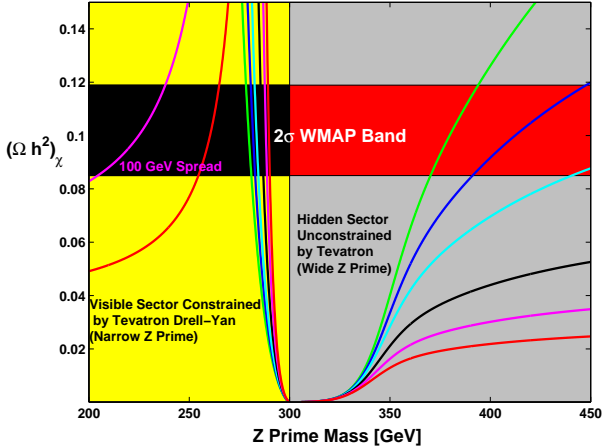}
\caption{An analysis of the  relic density of milli-charged
particles for the case  when kinetic mixing is included in the \st
$Z'$ model. The analysis is done for $M_{\chi} = $ 150 GeV,
$\bar\epsilon = .04$,
  and  $\delta = (.05, .075, .10, .15, .20, .25)$,
  where the values are in descending order for $M_{Z'}>300$ GeV.
  The red and black bands are the WMAP-3 constraints where the black
  band also produces an observable dilepton signal.
  The analysis shows that for $\bar {\epsilon}$ fixed,
 increasing $\delta$ increases the parameter space where the WMAP-3 relic density constraint is satisfied,
 while  allowing for  a detectable $Z$ prime signal as shown in Figs.(\ref{relic4},\ref{relic4a}).
}
\label{relic5}
\end{figure}

\section{Conclusion}
In the above we have given an analysis  of the \st extension of the
Standard Model with inclusion of the kinetic energy mixing in the
$U(1)_X\times U(1)_Y$ sectors. Such kinetic mixings are quite
generic in models with more than one $U(1)$ gauge group. It is shown
that in the model with both  the mass and kinetic mixing and in the
absence  of  matter in the hidden sector,  the sensitive parameter
which measures the deviation from the Standard Model is given by
$\bar\epsilon$ as defined by Eq. (\ref{barep}) which is a specific
combination of $\epsilon$ and $\delta$, where  $\epsilon$ measures
the mass mixing and $\delta$ measures the kinetic mixing. However,
when matter in the hidden sector is taken into account, electroweak
physics depends on both $\epsilon$ and $\delta$.   An analysis  of
the relic density of milli-charged dark matter which is generic in
\st extensions is given.  Here our analysis is in agreement with the
work of Ref. \cite{kctc} for the case when no kinetic mixing is
taken into account.  Inclusion of the kinetic mixing is seen to
increase the region of the parameter space where the relic density
constraints consistent with WMAP-3 can be satisfied.  We also
analyze the $Z'$ signal. As noted in Ref. \cite{kctc} on the branch
$M_{Z'} > 2M_{\chi}$ the dilepton signal from the $Z'$ decay is too
small to be  observed at colliders, and our results are in agreement
with this analysis. However, we note that on the branch $M_{Z'}<2
M_{\chi}$,  there  is a significant region of the parameter space
where the relic density constraints can be satisfied and the
dilepton signal from the $Z'$ decay via the Drell-Yan process  is
strong enough to be  observed at the Tevatron and at the LHC. The
analysis also shows that relic density constraints can be satisfied
for  values of $\bar\epsilon$ as low as $10^{-4}$ and even smaller
values are possible. An interesting issue concerns the detectability
of such dark matter in laboratory experiments which might put
further limits on the parameter space or on the component of the
relic density such matter can constitute. However, such an
investigation is outside the scope of this work.
\begin{center}
{\bf Acknowledgments}
\end{center}
 This work was supported in part by
the U.S. National Science Foundation under the grant
NSF-PHY-0546568.

\begin{widetext}
\section*{Appendix A: Details of the \st extension with kinetic mixing}

In this Appendix we give further details of the Stueckelberg extension with kinetic mixing.
 The mass matrix in the vector boson
sector,  after applying the transformation that diagonalizes the kinetic energy, is given by

 \be
 M^2=K^T M_{St}^2K,
 \ee
 and we display its explicit form below

\begin{equation}
M^2= \left(
\begin{array}{ccc}
        M_1^2\frac{(1-\epsilon\delta)^2}{1-\delta^2}+\frac{1}{4}\gamma^2 v^2\frac{\delta^2(1-2\epsilon\delta+\epsilon^2)}{1-\delta^2}
    &   M_1^2\frac{\epsilon (1-\epsilon\delta)}{\sqrt{1-\delta^2}}- \frac{1}{4} \gamma^2 v^2 \frac{\delta(1-2\epsilon\delta+\epsilon^2)}{\sqrt{1-\delta^2}}
    &   \frac{1}{4}  g_2 \gamma v^2 \frac{\delta\sqrt{1-2\epsilon\delta+\epsilon^2}}{\sqrt{1-\delta^2}}
    \\  M_1^2\frac{\epsilon (1-\epsilon\delta)}{\sqrt{1-\delta^2}}- \frac{1}{4} \gamma^2 v^2 \frac{\delta(1-2\epsilon\delta+\epsilon^2)}{\sqrt{1-\delta^2}}
    &   M_1^2 \epsilon^2 + \frac{1}{4} \gamma^2 v^2 (1-2\epsilon\delta+\epsilon^2)
    &   -\frac{1}{4} g_2 \gamma v^2 \sqrt{1-2\epsilon\delta+\epsilon^2}
    \\  \frac{1}{4}  g_2 \gamma v^2 \frac{\delta\sqrt{1-2\epsilon\delta+\epsilon^2}}{\sqrt{1-\delta^2}}
    &   -\frac{1}{4} g_2 \gamma v^2 \sqrt{1-2\epsilon\delta+\epsilon^2}
    &   \frac{1}{4}  g_2^2 v^2
\end{array}
\right). \label{msq}
\end{equation}
\end{widetext}
The eigenvalues of  Eq. (\ref{msq}) are
\begin{eqnarray}
     M_{\gamma}^2 = 0,\hspace{.25cm} M_{Z}^2 =(q-p)/2,\hspace{.25cm}  M_{Z'}^2  =(q+p)/2,
   \eeqn
   where
     \beqn
                          p    & = & \sqrt{\left( M_1^2\beta + \frac{\gamma^2 \beta +g_2^2}{4}v^2 \right) ^2-4 M_1^2 \frac{\gamma^2+g_2^2}{4}v^2
                          \beta},\nonumber \\
                           q     &  =  &   M_1^2\beta + \frac{\gamma^2 \beta
             +g_2^2}{4}v^2
\end{eqnarray}
and where    $ \beta = (1-2\epsilon\delta+\epsilon^2)/(1-\delta^2)$.
As expected in the limit $\epsilon=0=\delta$ the \st sector
decouples from the SM sector and one gets
\begin{eqnarray}
     M_{Z}  = \frac{\sqrt{g_2^2+\gamma^2}}{2}v,
    ~~M_{Z'}     = M_1.
\label{decoupled}
\end{eqnarray}
The expression for $M_Z$ of Eq. (\ref{decoupled}) is exactly the result  in the SM.
However,  quite remarkably  decoupling takes place also in the limit  when
$\epsilon$ and $\delta$ are non-vanishing but satisfy the relation $\epsilon = \delta$.
Here,
while the $\epsilon,  \delta$ dependence persists in the $M^2$ mass
matrix of Eq. (\ref{msq}), the eigenvalues are still given by Eq. (\ref{decoupled}) in this
limit.  Further, the coupling of the  $Z$  boson
to quarks and leptons reduces to that of the SM in this limit.
One may understand this  result  by examining the explicit form of the
transformation matrix  of Eq. (\ref{rotationmatrix}) which for the case $\epsilon=\delta$ becomes
\begin{equation}
T = \left(
\begin{array}{ccc}
        1
    &    \frac{\gamma}{\sqrt{g_2^2+\gamma^2}}\frac{\delta}{\sqrt{1-\delta^2}}
    &   -\frac{g_2}{\sqrt{g_2^2+\gamma^2}}\frac{\delta}{\sqrt{1-\delta^2}}
    \\  0
    &   -\frac{\gamma}{\sqrt{g_2^2+\gamma^2}}\frac{1}{\sqrt{1-\delta^2}}
    &   \frac{g_2}{\sqrt{g_2^2+\gamma^2}}\frac{1}{\sqrt{1-\delta^2}}
    \\  0
    &   \frac{g_2}{\sqrt{g_2^2+\gamma^2}}
    &   \frac{\gamma}{\sqrt{g_2^2+\gamma^2}}
\end{array}
\right).
\end{equation}
A direct substitution in the $Z$  and $Z'$  couplings then shows that they are
identical to the  case of the SM. These  analytic  forms explain the   limits  seen
numerically in the analysis of Fig.(\ref{200}).\\

Next we consider the case with arbitrary mass mixing and kinetic mixing.
Here before diagonalizing the mass matrix
it is convenient to perform the following orthogonal transformation $(O)$
given by
\begin{equation}
O= \left(
\begin{array}{ccc}
        \sqrt{1-\delta^2}    &    -\delta    &   0   \\
          \delta    &   \sqrt{1-\delta^2}    &  0    \\
            0    &  0    &   1
\end{array}
\right),
\end{equation}
which transforms the mass matrix to $\mathcal{M}^2=O^T M^{2}O$
\begin{equation}
\mathcal{M}^2 = \left(
\begin{array}{ccc}
        M_1^2        &               M_1^2 \bar\epsilon                          &           0 \\
        M_1^2 \bar\epsilon   &  M_1^2 \bar\epsilon^2 +\frac{v^2}{4}\gamma^2 (1+\bar\epsilon^2)  &   -\frac{v^2}{4}g_2\gamma \sqrt{1+\bar\epsilon^2} \\
        0            &       -\frac{v^2}{4}g_2\gamma \sqrt{1+\bar\epsilon^2}    &      \frac{v^2}{4}g_2^2
\end{array}
\right) \label{msq2}
\end{equation}
where $\bar \epsilon$ is defined so that
\begin{equation}
\bar\epsilon=\frac{\epsilon-\delta}{\sqrt{1-\delta^2}}.
\label{barep}
\end{equation}
We note that  the mass matrix Eq. (\ref{msq2}) looks exactly the
same as for the mass matrix one has  if there was just the \st mass
mixing except that  $\epsilon$ is  replaced by $\bar\epsilon$. Thus
the orthogonal matrix which diagonalizes Eq. (\ref{msq2}) will be
the same as for the StSM case except that the free parameter is
changed from $\epsilon$ to $\bar\epsilon$. We parameterize the
rotation matrix $\cal R$ defined by ${\cal R}^T {\cal
M}^2{\cal{R}}=Diag(M_{Z'}^2,M_{Z}^2,0)$ by the
following
\begin{widetext}
\begin{equation}
\label{R}{\cal R}= \left[\matrix{ \cos\psi
\cos\phi-\sin\theta\sin\phi\sin\psi & \sin\psi \cos\phi
+\sin\theta\sin\phi\cos\psi & -\cos\theta \sin\phi\cr
\cos\psi\sin\phi +\sin\theta\cos\phi\sin\psi & \sin\psi \sin\phi
-\sin\theta\cos\phi\cos\psi & \cos\theta \cos\phi\cr
-\cos\theta\sin\psi & \cos\theta\cos\psi & \sin\theta }\right],
\nonumber
\end{equation}
 where the angles are defined so that
\begin{eqnarray}
 \tan\theta &= & \frac{\gamma}{g_2},
 ~~~~\tan\phi   = \bar\epsilon ,
~~~~\tan 2\psi  = \frac{2\sin\theta
M_0^2\bar\epsilon}{M_1^2-M_0^2+(M_1^2+M_0^2-M_W^2)\bar\epsilon^2},
\end{eqnarray}
and $M_0=M_Z(\epsilon=\delta)=v\sqrt{g_2^2+\gamma^2}/2$,
$M_W=g_2v/2$. The transformation relating the initial basis and the
final diagonal basis is $V = [K(\delta) O(\delta){\cal{R}}(\bar
\epsilon)] E$, where $V^T= (C, B, A^3)$, and $E^T= (Z',
Z,A_{\gamma})$. The neutral current  interaction can now be  written
in the form \beqn
 \mathcal{L}_{NC} = J^T S(\bar \epsilon, \delta) {\cal{R}}(\bar\epsilon) E
\label{nc2}
\eeqn
where $J^T=(g_X J_X, \gamma J_Y, g_2 J_2^3)$,  and $S$ is given by
\begin{eqnarray}
 S(\bar \epsilon, \delta) =
\left[\matrix{1&-{\delta}/{\sqrt{1-\delta^2}}&0\cr
0&\sqrt{1+\bar\epsilon^2}&0\cr 0&0&1\cr}\right]\label{oneEW}.
\end{eqnarray}
When $J_X=0$, one finds that the neutral current
interaction of Eq. (\ref{nc2}) has no dependence on $\delta$.
Expressing the tree level interaction in terms of the reduced vector
and axial vector couplings as defined in Eq. (\ref{ncvs}), we find
\begin{eqnarray}
v_{f}   &=& \cos\psi
\left[\left(1-\bar\epsilon\sin\theta\tan\psi\right)T_f^3
                            -2\sin^2\theta\left(1-\bar\epsilon\csc\theta\tan\psi\right)Q_f\right],\\
 a_{f}   &=& \cos\psi \left[1-\bar\epsilon\sin\theta\tan\psi\right]T_f^3,\\
v'_{f}  &=& -\cos\psi
\left[\left(\tan\psi+\bar\epsilon\sin\theta\right)T_f^3
                            -2\sin^2\theta\left(\bar\epsilon\csc\theta+\tan\psi\right)Q_f\right],\\
 a'_{f}  &=& -\cos\psi \left[\tan\psi+\bar\epsilon\sin\theta\right]T_f^3.
\end{eqnarray}
Thus in the absence of the hidden sector matter, the effect on the
neutral current sector of including the kinetic mixing in the
\st model, is the replacement of $\epsilon$ by $\bar\epsilon$, and
in the limit $\bar{\epsilon} \to 0$ the  $Z$ couplings of the SM are
recovered. However,  with the  inclusion of matter in the hidden,
i.e., $J_X\neq 0$,  the neutral current sector will depend on both
$\bar\epsilon$ and $\delta$.  We  exhibit this below \beqn
 {\cal{L}}_{NC} =
 {\cal{L}}_{NC}(J_X=0) + g_X J_X^{\mu} \biggr[
\left({\cal{R}}_{11}-S_{\delta}{\cal{R}}_{21}\right)Z'_{\mu}
+\left({\cal{R}}_{12}-S_{\delta}{\cal{R}}_{22}\right)Z_{\mu}
+\left({\cal{R}}_{13}-S_{\delta}{\cal{R}}_{23}\right)A^{\gamma}_{\mu}\biggr]
\label{lnc}
   \eeqn\end{widetext}
where  $S_{\delta}$ appears in Eq.(\ref{lnc}), and is as defined in
Sec.(II). In this case  the decay modes of $Z'$ depend on both
$\epsilon$ and $\delta$ and an analysis of the branching ratios
should reveal the presence of kinetic mixing.
\\

 Further, as discussed in Sec.(IV) and  in Appendix D, the relic density analysis
depends sensitively on $\delta$.
\\\\\\

\section*{Appendix B: On the  origin of \\ milli-charged matter}
In this Appendix we illustrate the mechanism which generates the
milli-charge in the context of the analysis of this paper.  We start
with the kinetic mixing model \cite{holdom}
 with
two gauge fields $A_{1\mu}, A_{2\mu}$ corresponding to the gauge  groups $U(1)$ and $U(1)'$.
We choose  the following Lagrangian
$\mathcal{L} =\mathcal{L}_0 + \mathcal{L}_1$ where

\beqn \mathcal{L}_0 &=&
    - \frac{1}{4}F_{1\mu\nu}F_1^{\mu\nu}
    - \frac{1}{4}F_{2\mu\nu}F_2^{\mu\nu}
    - \frac{\delta}{2}F_{1\mu\nu}F_2^{\mu\nu},\nonumber\\
 \mathcal{L}_1 &=&
     J'_{\mu}A_1^{\mu}
    +J_{\mu}A_2^{\mu}.
       \label{onlykt}
\eeqn
To put  the kinetic energy term in its canonical form, one may
 use the transformation
\begin{equation}
 \left[\matrix{A_1^{\mu} \cr A_2^{\mu}}\right] \to K_0  \left[\matrix{A'^{\mu} \cr A^{\mu}}\right],
~~ K_0=\left[\matrix{\frac{1}{\sqrt{1-\delta^2}} & 0 \cr
\frac{-\delta}{\sqrt{1-\delta^2}} &1}\right].
 \label{ko}
 \end{equation}
However, the transformation that canonically diagonalizes the kinetic energy is not unique. Thus, for example,
   $K=K_0 R$ instead of $K_0$ would do  as well where $R$ is an orthogonal matrix
\begin{equation}
    R=\left[\matrix{ \cos\theta & -\sin\theta \cr \sin\theta &
    \cos\theta}\right].
\end{equation}
Here  $\mathcal{L}_1$ is given by
\begin{eqnarray}
 \mathcal{L}_1 =
     A'^{\mu}  \left[\frac{\cos\theta}{\sqrt{1-\delta^2}} J'_{\mu} +  \left(\sin\theta-\frac{\cos\theta\delta}{\sqrt{1-\delta^2}}\right) J_{\mu} \right]
       \nonumber\\ + A^{\mu}  \left[  -\frac{\sin\theta}{\sqrt{1-\delta^2}} J'_{\mu}  +
          \left( \cos\theta+\frac{\sin\theta\delta}{\sqrt{1-\delta^2}}\right)J_{\mu}
          \right].
   \end{eqnarray}
In this case we see that each of the massless states interacts with
the sources $J$ and $J'$.  However,  one may  choose $\theta$ to get
asymmetric solutions. For instance for the case $\theta =
\arctan\left[\delta/\sqrt{1-\delta^2}\right]$ one has

\begin{eqnarray}
 \mathcal{L}_1 =
     A^{\mu}  \left[\frac{1}{\sqrt{1-\delta^2}} J_{\mu}  -\frac{\delta}{\sqrt{1-\delta^2}} J'_{\mu} \right]
        + A'^{\mu} J'_{\mu}.
        \label{asymmetric}
   \end{eqnarray}
In this case while $A'$ interacts only with the source $J'$, $A$
interacts with both $J$ and $J'$, with the coupling to the source
$J'$ proportional to the kinetic mixing parameter $\delta$. We
identify $A$ with the physical photon field, $J$ with the physical
source arising from quarks and leptons, while  $A'$ is the
orthogonal massless state,
 and $J'$ is the source in the hidden sector.
 Here  the
coupling of the photon with the hidden sector  is  proportional to
$\delta$ and  thus   the hidden sector
is milli-charged if $\delta$ is small. \\

Next  we consider a model with kinetic mixing where a \st mechanism
generates  a mass term of the type considered in Eq. (1) \ \be
{\cal{L}}_{\rm{Mass}} = -\frac{1}{2} M_1^2  A_{1\mu}A_1^{\mu}
-\frac{1}{2} M_2^2 A_{2\mu}A_2^{\mu} - M_1M_2 A_{1\mu}A_2^{\mu}.
 \\\nonumber\ee In this case
diagonalizaton of  the mass  matrix fixes $\theta$ so that
\begin{equation}
\theta=\arctan\left[\frac{\epsilon\sqrt{1-\delta^2}}{1-\delta\epsilon}\right],
\end{equation}
and the   interaction Lagrangian is given by
\begin{eqnarray}
{\cal{L}}_{1} &= &\frac{1}{\sqrt{1-2\delta\epsilon+\epsilon^2}} \left( \frac{\epsilon-\delta}{\sqrt{1-\delta^2}} J_{\mu} +
       \frac{1-\delta\epsilon}{\sqrt{1-\delta^2}} J_{\mu}' \right) A_M^{\mu}
            \nonumber\\ & + &   \frac{1}{\sqrt{1-2\delta\epsilon+\epsilon^2}}
            \left(J_{\mu}- \epsilon J_{\mu}' \right)  A^{\mu}_{\gamma}.
\end{eqnarray}
\noindent Here for   the case  $\epsilon =0$  one finds that the
massless state, the photon $A_{\gamma}^{\mu}$, no longer couples
with the  hidden sector, while the massive mode $A_{M}^{\mu}$
couples with both the visible sector via $J$ and with the hidden
sector via $J'$. We  conclude, therefore, that in the absence of the
Stueckelberg mass mixing, for the case when only one mode is
massless, there are no milli-charged particles coupled to the photon
field.   Thus milli-charge couplings appear in this case only when
the Stueckelberg mixing parameter $\epsilon$ is introduced.  Thus
for the case when only one mode is massless the kinetic mixing by
itself does  not allow milli-charges but the Stueckelberg mass
mixing model does.

\section*{Appendix C:
Analysis of relic density of milli-charged dark matter in the hidden
sector} The analysis of relic density involves the integral of the
thermally averaged  cross  section from the current temperature to
the freeze out  such that \cite{Nath:1992ty} \beqn J(x_f) =
\int_0^{x_f}\langle \sigma v \rangle dx .\eeqn Here $\sigma$ is the
annihilation cross section for the process  $\chi \bar\chi \to f\bar
f$ where $f$  denotes a quark or a lepton, and $v$ is the relative
velocity of the annihilating  Dirac fermions  $\chi$ and $\bar
\chi$. We use the notation, $x=T/M_{\chi}$ in units where the
Boltzman constant is unity, and $x_f$ is the value of $x$ at the
freeze out temperature. The thermally averaged cross section
$\langle \sigma v \rangle$ is given by
 \be
\langle \sigma v \rangle  = \frac{\int_0^{\infty} dv v^2   (\sigma v)  e^{-(v^2/4x)}}
  {\int_0^{\infty} dv v^2 e^{-(v^2/4x)}}.
 \label{sigv}
  \ee
Using the analysis of Ref. \cite{kctc}  $(\sigma v)$ is given by
\beqn
(\sigma v)   \simeq  f_1(s)({\xi}^2_L + {\xi}^2_R) +f_2(s)({\xi}_L {\xi}_R)
\label{sigv1}
 \eeqn
 where the functions $f_1$ and $f_2$  depend on the square of the CM energy $s$ and are given by
\beqn  f_1(s) &= & C_f v \frac{\beta_{f}}{s
\beta_{\chi}}[s^2(1+\frac{1}{3}\beta^2_{f}\beta^2_{\chi})+4
M^2_{\chi}(s-2m^2_f)]\nonumber \\
f_2(s)&=& C_f v \frac{\beta_{f}}{s \beta_{\chi}}[8 m^2_f(s+2
M^2_{\chi})] \label{sigv2} .\eeqn Here $C_f = N_f/(32 \pi)$, $N_f =
(1,3)$ for $f=(\rm lepton, \rm quark)$, $\xi_L$ and $\xi_R$ are  as
defined in Ref. \cite{kctc} and contain the photon, the $Z$  and the
$Z'$ poles, the latter augmented with a Breit-Wigner distribution
for thermal averaging, as the largest contribution to $\langle
\sigma v \rangle$  for $4 M^2_{\chi} \gg M^2_{Z}$ arises  from
integration over the $Z'$ pole. In the vicinity of the $Z'$ pole we
can expand $(\sigma v)$ as follows \be (\sigma v) \simeq  C_{1} +
C_{2}v^2 + \frac{C_{3}M^{-4}_{\chi} +C_{4}M^{-2}_{\chi} v^2}{( v^2 -
\epsilon_R )^2 +  \gamma^2_R } \ee where $\epsilon_R$ and $\gamma_R$
are given by \be \epsilon_R= (M_{Z'}^2-4M_{\chi}^2)/M_{\chi}^2~, ~~~
\gamma_R =\Gamma_{Z'} M_{Z'}/M_{\chi}^2 \ee and where  $C_i (i=1-4)$
can be read off from the expansion of $s$ in Eq. (\ref{sigv1}) and
Eq. (\ref{sigv2}).
  Using the technique of Ref. \cite{Nath:1992ty}  the integral  $J(x_f)$ is given by
  \begin{widetext}
\be J(x_f)= x_{f} C_1+3 x^2_{f}C_2 + \frac{1}{2\sqrt{\pi}
M^4_{\chi}}
\biggr[\frac{C_{3}}{{\epsilon}_{R}}I_{2}+M^2_{\chi}C_{4}(I_{1}+
I_{2})\biggr] \ee where \be I_{1}= \frac{1}{2}\int_0^{\infty} dy
y^{-1/2}e^{-y} Ln\biggr[\frac{(4 x_{f}y-\epsilon_{R})^2
+\gamma^2_{R}}{\epsilon^2_R +\gamma^2_R}\biggr],\ee
\be
I_{2}=\frac{\epsilon_{R}}{\gamma_{R}}\int_0^{\infty} dy
y^{-1/2}e^{-y} \biggr[\arctan\biggr[\frac{4
x_{f}y-\epsilon_{R}}{\gamma_{R}}\biggr] + \arctan
\biggr[\frac{\epsilon_{R}}{\gamma_R}\biggr]\biggr]. \ee From the
above  one then obtains  the relic density using the standard
relations as discussed, for example, in Ref. \cite{Nath:1992ty}.
\end{widetext}

 \begin{widetext}
\section*{Appendix D:  The effect of $\delta$ on the cross section}

\noindent
The kinetic mixing parameter
$\delta$ enters  Eq. (\ref{lnc}) via the hidden sector where
\begin{eqnarray}
\mathcal{L}_{NC}^{\rm{hid}} &=& \bar{\chi}\gamma^{\mu}
                        \left[ \epsilon_{Z'}^{\chi} Z'_{\mu}+\epsilon_{Z}^{\chi} Z_{\mu}+\epsilon_{\gamma}^{\chi}
                        A^{\gamma}_{\mu}\right]\chi.
\end{eqnarray}
In the limit $\bar\epsilon, \delta \ll 1$,  the  couplings to leading order  are

\begin{eqnarray}
    \epsilon_{Z'}^{\chi}        &=&     g_X Q_X \left[  {\cal{R}}_{11}-\frac{\delta}{\sqrt{1-\delta^2}}{\cal{R}}_{21} \right]
                                   \simeq      g_X Q_X,\\
    \epsilon_{Z}^{\chi}         &=&     g_X Q_X \left[  {\cal{R}}_{12}-\frac{\delta}{\sqrt{1-\delta^2}}{\cal{R}}_{22} \right]
                                    \simeq     \bar\epsilon g_X Q_X \sin \theta  \left[  1+\frac{\delta}{\bar\epsilon} \right],\\
    \epsilon_{\gamma}^{\chi}    &=&     g_X Q_X \left[  {\cal{R}}_{13}-\frac{\delta}{\sqrt{1-\delta^2}}{\cal{R}}_{23}  \right]
                                    \simeq     -\bar\epsilon g_X Q_X \cos \theta  \left[  1+\frac{\delta}{\bar\epsilon} \right].
\end{eqnarray}
\\
For the purpose of illustration in this Appendix, and only here, we will work in the limit $m_f \sim 0$ , $v
\sim 0$ and $\Gamma_{Z'}\sim 0$ so that
\\
\begin{equation}
\sigma v \simeq \frac{M_\chi^2N_f}{2 \pi}\biggr [
\xi_L^2+\xi_R^2\biggr ],
\end{equation}
where for $4M_{\chi}^2 \gg M_Z^2$,
\begin{eqnarray}
  \xi_{L,R} &\simeq & \frac{\epsilon_\gamma^\chi e Q_{\rm em}^f }{4 M_\chi^2}
           +\frac{\epsilon_Z^\chi \epsilon^{f_{L,R}}_Z }{4 M_\chi^2- M_Z^2}
           +\frac{\epsilon_{Z'}^\chi \epsilon^{f_{L,R}}_{Z'}}{4
           M_\chi^2-M_{Z'}^2}\nonumber \\
           &=&\alpha_{L,R}+\frac{\beta_{L,R}}{4
           M_\chi^2-M_{Z'}^2}.
\end{eqnarray}
\newline
Defining $x=(4M_{\chi}^2-M_{Z'}^2)$,  $\sigma v$  can
be further simplified as
\begin{eqnarray}\label{xsection}
\sigma v &\simeq  & a+\frac{2b}{x}+\frac{c}{x^2},
\end{eqnarray}
where $a,b,c$ are functions of $\eta =(1+\delta/\bar\epsilon)$ and are given by
\begin{eqnarray}
a &=& \frac{M_\chi^2N_f}{2 \pi} (\alpha_L^2+\alpha_R^2)\nonumber\\
b &=& \frac{M_\chi^2N_f}{2 \pi}(\alpha_L\beta_L+\alpha_R\beta_R),\nonumber\\
 c &=& \frac{M_\chi^2N_f}{2 \pi} (\beta_L^2+\beta_R^2).
\end{eqnarray}
The $\eta$ dependence can then be read off so that
$a(\eta)=\eta^2 a(1),  b(\eta)=\eta b(1),
c(\eta)=c(1)$.
The cross section described by Eq. (\ref{xsection}) has a dip when
 $x_{min}=-c/b$ which gives $(\sigma v)_{min} = a-b^2/c$.
Now $x_{min}$ and $(\sigma v)_{min}$  have an  $\eta$ dependence
of the form $x_{min}(\eta)=x_{min}(1)/\eta$, and $(\sigma v)_{min} (\eta)=
\eta^2(\sigma v)_{min} (1) $ which implies  that for  the case $\eta>1$  a larger  $\eta$
shrinks $x_{min}$ and moves the dip closer to the resonance point.  At the same time it also
increases $(\sigma v)_{min}$.   More interestingly, a negative $\eta$
will switch the dip to the other side of the resonance. These effects are illustrated in
the analysis of Fig.(\ref{J2}).
The analysis above explains why the inclusion of kinetic mixing enhances the region where
the relic density constraint is satisfied.\\
 \end{widetext}

\end{document}